\documentclass{pasj01}
\draft 
\Received{$\langle$reception date$\rangle$}
\Accepted{$\langle$acception date$\rangle$}
\Published{$\langle$publication date$\rangle$}

\begin{document}

\title{Discovery of  annular X-ray emission centered on MAXI\,J1421$-$613: Dust-scattering X-rays?}
\author{Kumiko K.  \textsc{Nobukawa}\altaffilmark{1}, Masayoshi  \textsc{Nobukawa}\altaffilmark{2}, and Shigeo  \textsc{Yamauchi}\altaffilmark{1}}
\altaffiltext{1}{Department of Physics, Faculty of Science, Nara Women's University, Kitauoyanishi-machi, Nara, Nara 630-8506, Japan}
\altaffiltext{2}{Faculty of Education, Nara University of Education, Takabatake-cho, Nara, Nara 630-8528, Japan}
\email{kumiko@cc.nara-wu.ac.jp}


\maketitle

\begin{abstract}
We report the discovery of an annular emission of $\sim$\timeform{3'}--\timeform{9'} radius around the center of a transient source, an X-ray burster MAXI J1421$-$613, in the Suzaku follow-up analysis. The spectrum of the annular emission shows no significant emission-line structure, and is well explained by an absorbed power law model with a photon index of $\sim$4.2. These features exclude the possibility that the annular emission is a shell-like component of a supernova remnant.  The spectral shape, the time history, and the X-ray flux of the annular emission agree with the scenario that the emission is due to a dust-scattering echo. The annular emission is made under a rare condition of the dust-scattering echo, where the central X-ray source, MAXI J1421$-$613, exhibits a short time outburst with three X-ray bursts and immediately re-enters a long quiescent period. The distribution of the hydrogen column density along the annular emission follows that of the CO intensity, which means that MAXI J1421$-$613 is located behind the CO cloud. We estimate the distance to MAXI J1421$-$613 to be $\sim$3~kpc assuming that the dust layer responsible for the annular emission is located at the same position as the CO cloud.
\end{abstract}

\section{Introduction}
X-rays from an X-ray source are scattered by interstellar dust in the line of sight \citep{Overbeck65}. A dust-scattering halo is observed surrounding a bright  X-ray source which is located behind a large amount of dusts. 
The dust-scattering halos provide information on the interstellar dust, such as the size distribution and composition of the grains (e.g., \cite{Predehl95}, \cite{Draine03}).  
By measuring the temporal variation of a halo,  \citet{Predehl00} constrained a line of sight position of Cygnus X-3. 

If the X-ray emission with a short duration time is scattered by dust clouds, ring echoes are detected. Such samples have been detected around several gamma-ray bursts (e.g., \cite{Vaughan04, Vaughan06, Tiengo06, Vianello07, Pintore17}) and Galactic sources (e.g., \cite{Svirski11, Mao14}). Only three Galactic samples  have  clear, well-defined, and  large rings with a several arcminute scale (e.g., 1E\,1547.0$-$5408, Circinus X-1, and V404 Cygni; \cite{Tiengo10, Heinz15, Vasilo16, Heinz16}).  Most samples are accompanied with multiple rings, and each ring corresponds to a different dust layer or a different outburst. The thickness of the observed rings are different according to the situation; they reflect the duration time of the burst, the thickness of the dust layer,  and the spatial resolution of the observation instrument. 

An outburst of the new X-ray source MAXI\,J1421$-$613 was detected by the MAXI Nova Alert System  \citep{Negoro10} on 2014 January 9 \citep{Morooka14}. 
During the outburst, MAXI\,J1421$-$613 showed at least three type I X-ray bursts, and thus it is categorized into a neutron-star low-mass X-ray binary \citep{Bozzo14, Serino15}. 
The first burst with the observed flux of $1.7\times10^{-9}$~erg~cm$^{-2}$~s$^{-1}$ in the 3--10~keV band was detected on January 10 and lasted about 20~s \citep{Bozzo14}.  
The second one occurred on January 16 is the brightest among the three. The bolometric flux was calculated to be $7\times10^{-8}$~erg~cm$^{-2}$~s$^{-1}$  and the duration time was about 36~s \citep{Serino15}. The third one observed on January 18  has the bolometric flux of $3.2\times10^{-8}$~erg~cm$^{-2}$~s$^{-1}$ and the duration time of about 40~s \citep{Serino15}.  

After the MAXI alert, the Swift  X-Ray Telescope (XRT) began a target-of-opportunity observation on the same day and observed the source approximately every 2 days until February 3. Suzaku performed a follow-up observation since January 31, and the observation continued for three days. But it did not detect the source and thus  the 3$\sigma$ upper limit of the source flux of $1.2\times10^{-13}$~erg~cm$^{-2}$~s$^{-1}$ (0.5--10~keV) was obtained \citep{Serino15}.  Chandra also observed MAXI\,J1421$-$613 on February 8 with the effective time of 969~s.  No source was detected significantly, and the 95\%-confidence upper limit on the flux was measured  to be $8.1\times10^{-14}$ ~erg~cm$^{-2}$~s$^{-1}$ (0.2--10 keV; \cite{Chakrabarty14}). 

\citet{Serino15} analyzed the spectra during the outburst by utilizing the MAXI Gas Slit Camera (GSC) and the Swift XRT follow-up observations. The authors revealed that the spectra during the outburst can be explained by thermal Comptonization of multi-color disk blackbody emission.  The photon index $\Gamma \sim2$ is a typical for low-mass X-ray binaries and remains almost constant during the outburst. 
Assuming the empirical maximum luminosity of X-ray bursts, \citet{Serino15} estimated the maximum distance to be 7~kpc from the observed peak flux.   

In this paper, we report on the discovery of an annular X-ray emission around MAXI\,J1421$-$613 from the Suzaku data. Based on spectral and radial profile analyses of the annular emission,  we discuss its possible origin: (1) a supernova remnant  or (2) a dust-scattering echo.

\section{Observations and data reduction}
We utilized the X-ray Imaging Spectrometer (XIS) data \citep{Koyama07} aboard Suzaku \citep{Mitsuda07}. 
The XIS consists of four X-ray CCD cameras, each placed on the focal plane of the X-Ray Telescope (XRT; \cite{Serle07}). 
A field of view (FOV) of the XIS is $\timeform{17.8'}\times\timeform{17.8'}$. Three of the sensors (XIS0, 2, and 3) employ front-illuminated (FI) CCDs, while the other (XIS1) has a back-illuminated (BI) CCD. The entire region of XIS2 and one-fourth of XIS0 have been out of function since 2006 November and 2009 June, respectively. 

We analyzed  data with the analysis software package HEAsoft 6.22.1 and the Suzaku calibration database (CALDB) released in 2016 February. The spectral analysis was performed with XSPEC 12.9.1.  The data were screened by the standard event selection criteria for the XIS data processing. 
The response file (arf) and redistribution file (rmf) were produced by {\tt xissimarfgen} and {\tt xisrmfgen} \citep{Ishisaki07}, respectively. 
The non-X-ray background (NXB) was estimated by {\tt xisnxbgen} \citep{Tawa08} and was subtracted from spectra and images in the following analysis.
 For the spectral analysis, we used nearby blank-sky data as background.  The observation log is summarized in table~\ref{obs_log}. 
 Throughout the paper,  FI and BI spectra were fitted simultaneously, but only the FI spectra (XIS0$+$3) are displayed in the figures for brevity. Error bars given in figures show the 1$\sigma$ statistical errors.

\begin{table*}
  \tbl{Observation log.}{%
  \begin{tabular}{lcccccc}
  \hline
  Object  Name 	& Obs.ID & \multicolumn{2}{c}{Pointing direction} 			& \multicolumn{2}{c}{Observation date (UT)} & Exposure (ks) \\
              		 &             & $\alpha_{\rm J2000.0}$ (\timeform{D}) 	& $\delta_{\rm J2000.0}$ (\timeform{D})  &  Start 	&  End		    &                             \\  
  \hline
 MAXI\,J1421$-$613 & 908003010 & 215.40 &  $-$61.61  &  2014-01-31 12:20:40 & 2014-02-03 14:00:17 & 48.8 \\
 ASO0304 (background) &  504054010 & 213.34 & $-$62.08  & 2009-07-24 21:42:28 & 2009-07-26 03:29:22  & 44.2 \\
  \hline
  \end{tabular}}\label{obs_log}
 \end{table*}

\begin{figure*}
 \begin{center}
\includegraphics[width=16cm]{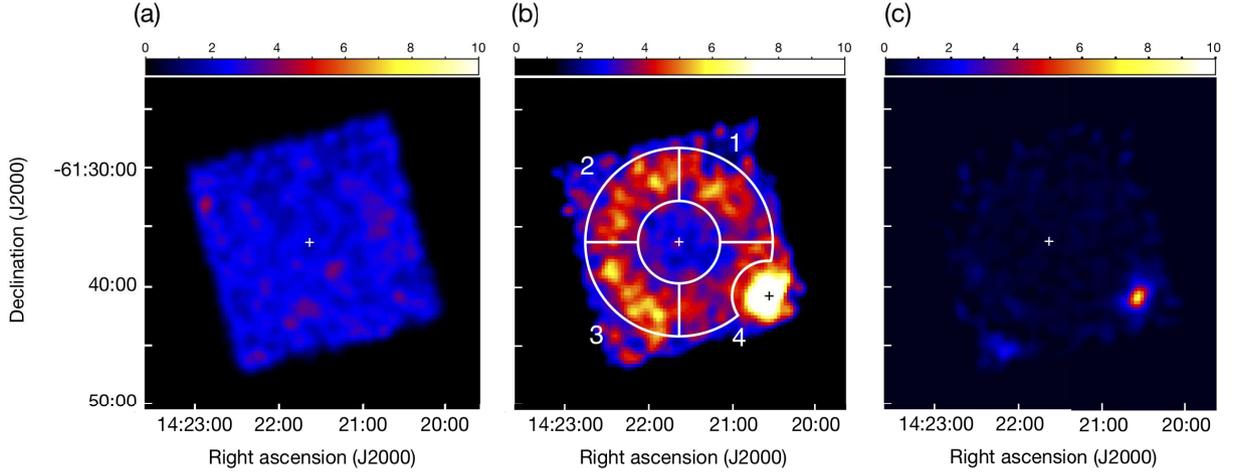}
 \end{center}
 \caption{X-ray images of the XIS FOV containing MAXI\,J1421$-$613 in the 0.5--2~keV (a), 2--5~keV (b), and 5--8~keV bands (c). The color scales are shown in the arbitrary unit. A white cross in (b) shows the position of MAXI\,J1421$-$613, while a black cross shows Suzaku\,J1420.5$-$6141 (see text). We extracted a source spectrum from the whole annular region surrounded by the white lines. We divided the annular region into the four parts (1--4) and also performed spectral analysis for each quadrant.}\label{3band_image}
\end{figure*}

\section{Analysis and Results}\label{analysis}

The X-ray images of the entire FOV of Suzaku XIS in the 0.5--2 keV, 2--5 keV and 5--8 keV bands are given in figure~\ref{3band_image}, where the position of MAXI\,J1421$-$613 is shown by the white cross. No significant X-ray emission was found at this position. 
Another point source was found (black cross in figure~\ref{3band_image}) at the position of $(\alpha, \delta)_{\rm J2000.0} = (\timeform{14h20m34s.07},\timeform{-61D41'00''.33})$ with the positional uncertainty of $\timeform{19"}$ \citep{Uchiyama08b}. No catalogued source has been reported so far at this position, hence is newly named as Suzaku\,J1420.5$-$6141.  The spectrum of Suzaku\,J1420.5$-$6141 was made from a circle with a radius of $\timeform{3'}$. The best-fit power-law model gave absorption column density, photon index, and observed flux in the 2--10~keV band to be $N_{\rm H} = 4.0^{+0.9}_{-0.8}\times10^{22}$~cm$^{-2}$, $\Gamma=1.6\pm0.3$, and $F_{\rm{2-10~keV}}=(1.4\pm0.1)\times10^{-12}$~ergs~cm$^{-2}$~s$^{-1}$, respectively. No time variability within $<10$\% was found.

\begin{figure}
 \begin{center}
\includegraphics[width=10cm]{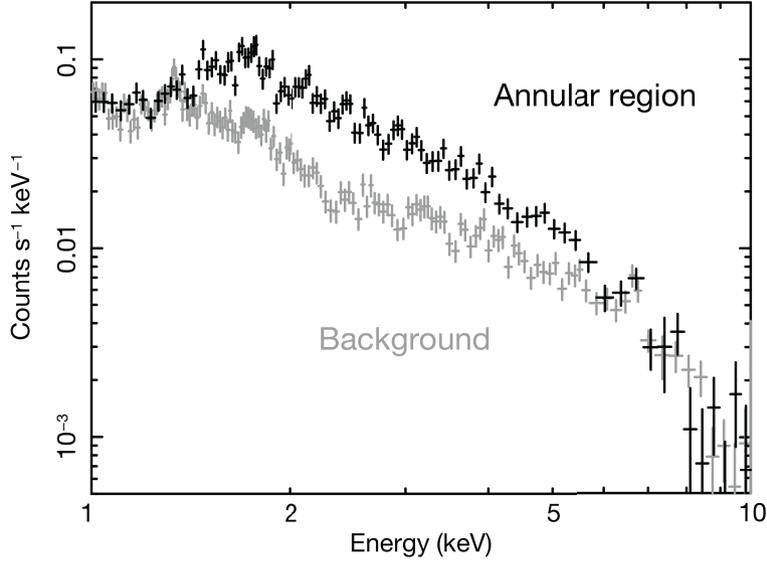}
 \end{center}
 \caption{Source spectrum (black) extracted from the whole annular region indicated by white lines in figure~\ref{3band_image}(b) and background spectrum (gray) extracted from the blank-sky data. }\label{spec_src_bgd}
\end{figure}

We found clear an annular emission around MAXI\,J1412$-$613. 
The width is about $\timeform{6'}$ and the flux is reasonably high with uniform distribution except Suzaku\,J1420.5$-$6141.
The spectrum was made from the annular region excluding Suzaku\,J1420.5$-$6141. The background spectrum was obtained from the nearby position (table~\ref{obs_log}). These spectra are shown in figure~\ref{spec_src_bgd}. The background subtracted spectrum was fitted with either a phenomenological power-law model or an optically thin thermal plasma model.  The best-fit results are given in  table~\ref{fit_param}. 
The hydrogen column density obtained by the power-law model is  $N_{\rm H} = (3.7\pm0.5)\times10^{22}$~cm$^{-2}$, which is consistent with that of MAXI\,J1412$-$613 measured by \citet{Serino15}, $N_{\rm H} = 4.8^{+1.3}_{-1.1}\times10^{22}$~cm$^{-2}$. 

\begin{table}
  \tbl{Best-fit parameters of the annular emission spectrum$^{\ast}$. }{%
  \begin{tabular}{lcc}
  \hline
Parameter	  						& phabs$\times$apec	&  phabs$\times$power-law	 		\\	
 \hline
$N_{\rm H} (10^{22}$~cm$^{-2}$)		& $2.3\pm0.3$         	    	& $3.7\pm0.5$			 \\ 
$\Gamma$						&	--				& $4.2\pm0.3$					 \\
$kT$ (keV)						& $1.4\pm0.2$			&	--						 \\
Abundance (solar)					& $<0.09$				&	--						 \\
Flux$^{\dag}$						& $1.9\pm0.1$			& $2.5\pm0.1$					 \\
  \hline 
 reduced $\chi^2$ (d.o.f.)				& 1.17  (112)			& 1.09 (113)					 \\
   \hline
   \multicolumn{3}{l}{$^{\ast}$ Errors are quoted at the 90\% confidence levels.}\\
   \multicolumn{3}{l}{$^{\dag}$ Unabsorbed values in the 2--5~keV band in unit of $10^{-12}$~erg~cm$^{-2}$~s$^{-1}$. }\\
  \end{tabular}}\label{fit_param}
 \end{table}

\begin{figure}
\begin{center}
\includegraphics[width=8cm]{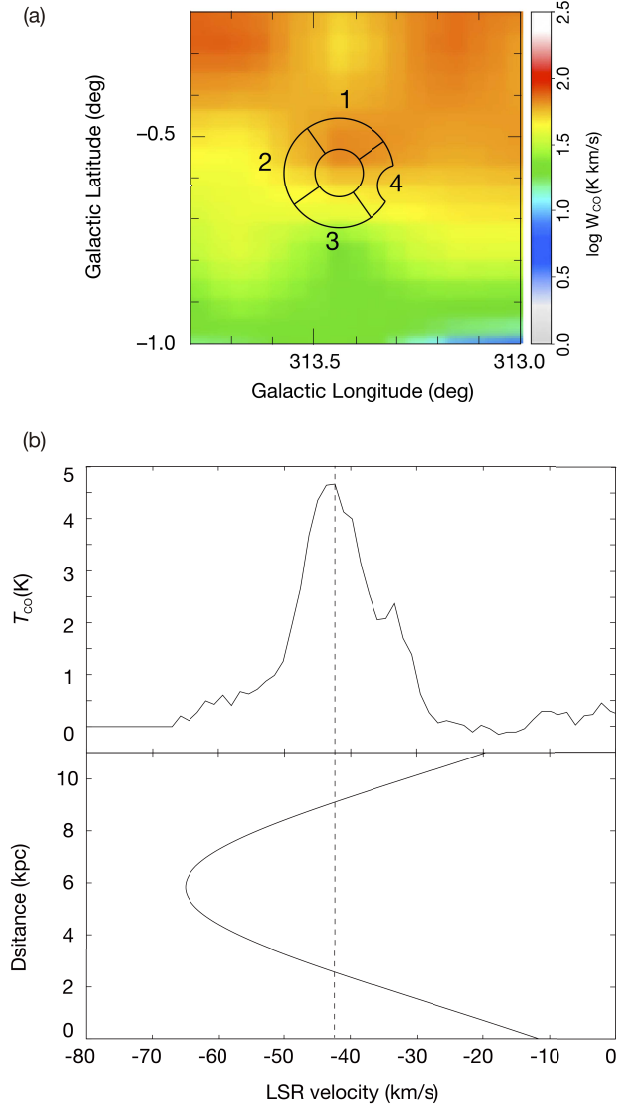}
 \end{center}
 \caption{(a) CO map around MAXI\,J1421$-$613 \citep{Dame01}.  The quadrants of the annular region are overlaid. (b) (Top) CO spectrum in the direction of the quadrant~1 taken from \citet{Dame01}. Dashed line indicates the spectral peak at $\sim43$~km~s$^{-1}$. (Bottom) Correlation between distance from the sun and radial velocity (LSR) in a direction of quadrant~1 ($l=\timeform{313.D4}$) inferred from the Galactic rotation curve obtained by \citet{McClure07}. }\label{fig:CO}
\end{figure}

The region of the annular emission is shown on the flux map of CO \citep{Dame01} in figure~\ref{fig:CO}. Referring the $N_{\rm H}$ distribution, we divided the region into four quadrants (quadrant 1--4). The spectrum of each quadrant was fitted with a phenomenological power-law model. The photon indices of the four parts are consistent with the best-fit value of the whole annular emission ($\Gamma = 4.2$) within the 90\% confidence levels. 
We fixed the photon index to $\Gamma = 4.2$ and made other parameters free. The best-fit $N_{\rm H}$ values are $4.5\pm0.4$,  $3.7\pm0.4$, $3.1\pm0.3$, and $3.2\pm0.4$ in the unit of $10^{22}$~cm$^{-2}$ for the quadrant 1, 2, 3, and 4, respectively. 
The largest $N_{\rm H}$ at the quadrant 1 corresponds to the CO peak, while the smallest $N_{\rm H}$ at quadrant~3 is out of the CO peak \citep{Dame01}.  

In order to give a constraint on the cloud position, adopting the Galactic rotation curve by \citet{McClure07} of $R_{\rm o}=8.5$~kpc and $V_{\rm o}=220$~km~s$^{-1}$, we plotted the correlation map of the LSR velocity and the distance from the sun (see figure~\ref{fig:CO}). The LSR of $\sim43$~km~s$^{-1}$ gives two solutions for the distance of the cloud of $\sim2.6$~kpc and $\sim9$~kpc.

We made four annular spectra of \timeform{3.'5}--\timeform{5'}, \timeform{5'}--\timeform{6'}, \timeform{6'}--\timeform{7'}, 
and \timeform{7'}--\timeform{8'} and investigated spectral variation with a radius.  We fitted the spectra with the phenomenological power-law model with fixing the $N_{\rm H}$ value of  3.7$\times$10$^{22}$~cm$^{-2}$. No clear differences in the spectral shape were found.

We also extracted spectra of the annular emission from the three periods and examine the spectral change with time. The periods are named as period 1, 2 and 3 (see table~\ref{fit_param2}). 
The spectrum was fitted with the phenomenological power-law model with fixing the $N_{\rm H}$ value of  3.7$\times$10$^{22}$~cm$^{-2}$. 
The best-fit spectral parameters are listed in table~\ref{fit_param2}. 

\begin{table*}
  \tbl{Best-fit parameters of each period spectrum$^{\ast}$. }{%
  \begin{tabular}{lccc}
  \hline
								& period 1				& period 2				& period 3	\\		
Start time (UT)						& 2014-01-31 12:20:40	& 2014-02-01 13:47:37	& 2014-02-02 15:37:31\\
Exposure time (ks)					& 17.1				&15.2				&16.4		\\
\hline
Parameter	  						&\multicolumn{3}{c}{phabs$\times$power-law}	\\	\hline
$N_{\rm H} (10^{22}$~cm$^{-2}$)		& 3.7 (fixed)			& 3.7 (fixed)			& 3.7 (fixed)	 \\ 
$\Gamma$						& $4.2\pm0.2$			& $4.3\pm0.2$			& $4.1\pm0.2$		 \\
Flux$^{\dag}$						& $2.6\pm0.1$			& $2.4\pm0.1$			& $2.2\pm0.1$		 \\
  \hline 
 reduced $\chi^2$ (d.o.f.)				& 1.11 (79)			& 1.24 (69)			& 0.96 (69)		 \\
   \hline
   \multicolumn{3}{l}{$^{\ast}$ Errors are quoted at the 90\% confidence levels.}\\
   \multicolumn{3}{l}{$^{\dag}$ Unabsorbed values in the 2--5~keV band in unit of $10^{-12}$~erg~cm$^{-2}$~s$^{-1}$.} \\
  \end{tabular}}\label{fit_param2}
 \end{table*}

In order to examine the expansion of the annular emission, we made a radial profile from three different periods of the observation and fit the radial profiles with a gaussian. 
The gaussian centroid of period 1, 2, and 3 was measured to be $\timeform{5.'4\pm0.'2}$, $\timeform{5.'9}\pm{0.'2}$, and $\timeform{5.'9\pm0.'3}$, respectively.

We also checked whether the annular emission is detected by other satellites. 
Swift performed two observations of MAXI\,J1421$-$613 with the Photon Counting (PC) mode in almost the same period as the Suzaku observations;  one observation started on January 30 (Obs ID $=$ 00033098009) and the other one started on February 1 (Obs ID $=$ 00033098010).  The total exposure time of the two observations is 2.4 ks. 
We extracted a radial profile centered on the position of MAXI\,J1421$-$613 in the 1--5~keV band using the Swift data. The number of counts in each radial bin are divided by the corresponding area. We also divided the profile by the effective exposure time derived from the exposure map in order to take in to account telescope vignetting, the CCD bad pixels and columns, and the attitude variations. We found that the profile has a hint of a peak at $\sim\timeform{6'}$. The count rate of the annular emission is measured to be $0.03$--$0.05$~counts~s$^{-1}$, which approximately corresponds to the flux of $(2$--$3)\times 10^{-12}$~erg~cm$^{-2}$~s$^{-1}$. These values are consistent with the radius and the flux of the annular emission observed by Suzaku. However, the swift data could not constrain the spectral parameters due to insufficient statistics.
We also checked  the Chandra/ACIS-S data, but we cannot distinguish the annular emission  due to a short exposure time (1 ks) and the low net counts of 68 within the whole FOV (the 2--5 keV band).

\section{Discussion}
\subsection{Origin of the annular emission}
We found an annular X-ray  emission around the low-mass X-ray binary MAXI\,J1421$-$613 from the Suzaku data. We discuss its origin based on the results. 

Composite type SNRs have a shock-heated thermal shell with a Compact Central Object (CCO) (e.g., CTB109, RCW103;  \cite{Sasaki04, Tuohy80}). The observed morphology indicates one possibility that the annular emission is the shock-heated shell of SNR with a CCO (MAXI\,J1421$-$613) in the center, where the CCO would be a neutron star of the low mass X-ray binary (X-ray burster).  The flux and power-law index ($\Gamma$) of MAXI\,J1421$-$613 during the outburst phase are $10^{-9}$--$10^{-10}$~erg~cm$^{-2}$~s$^{-1}$ and $\Gamma \sim2.1$, respectively \citep{Serino15}. The value of $\Gamma$ is consistent with a disk-blackbody model for a neutron star low-mass binary \citep{Mitsuda84}.  

The spectrum of the annular emission was nicely fitted with a power-law model with $\chi^2$/d.o.f.$ = 1.09$. The spectrum shows no prominent line, setting the abundance upper-limit of $<0.09$ solar (table~\ref{fit_param}). Such low abundance is unreasonable for the ejecta or interstellar gas of SNRs. No radio shell has been reported. 
Thus, the annular emission would not be due to a thermal plasma in a composite type SNR.  
The power-law index of $\Gamma\sim4.2$ is out of the standard synchrotron radiation found in a shell region of SNRs ($\Gamma \sim$2--3, e.g., SN1006: \cite{Bamba2003}; RX J1713$-$3946: \cite{Koyama1997}).  The mean flux of  the annular emission, $(2.5\pm0.1)\times10^{-12}$~erg~cm$^{-2}$~s$^{-1}$ (table~\ref{fit_param}), is less than the peak flux of MAXI\,J1421$-$613. These facts support that the annular emission is made by dust scattering.

The dust scattering echo has a halo-like profile in persistent bright point sources (e.g. many Galactic binaries, \cite{Predehl95}), while a ring-like structure with many concentric sub-rings is found in highly variable X-ray sources, i.e. 1E\,1547.0$-$5408, Circinus~X-1, and V404~Cygni \citep{Tiengo10, Xiang11, Heinz15, Heinz16, Vasilo16}.  
The differential cross section of X-ray dust scattering depends on the photon energy and is given by  $E^{-\Gamma}$  in the optically thin limit where one can approximate $(1 - \rm{e}^{-\tau})$ as $\tau$, where $\tau$ is the opacity. The index $\Gamma$ is about 2 \citep{Draine03}.  The observed $\Gamma$ of the annular emission is $\sim4.2$ which is larger than MAXI\,J1421$-$613 of $\Gamma\sim2.1$ by $\Delta\Gamma\sim2$ \citep{Serino15}.  
This value strongly supports the dust scattering echo scenario. 

If the annular emission is dust scattering of the short (but not instantaneous) outburst, the radius may expand with time. 
In fact, we found a hint of the radius expansion from $\sim$\timeform{5.'4} (period 1) to $\sim$\timeform{5.'9} (periods 2 and 3). 
We also found a hint of flux decrease from 2.6$\times$10$^{-11}$~erg~cm$^{-2}$~s$^{-1}$ (period 1) to 2.2$\times$10$^{-11}$~erg~cm$^{-2}$~s$^{-1}$ (period 3) without the spectral change. 
In the slight expansion of $\sim$\timeform{30''}, the differential cross section of dust scattering shows a little decrease \citep{Draine03}. 
The time history is well consistent with those expected from the dust scattering echo scenario.

\subsection{Location of MAXI J1421$-$613 and dust layer}

In this section, we estimate the distance to MAXI J1421$-$613 and a dust layer. 

MAXI J1421$-$613 entered into an outburst phase around 2014 January 7 after a long period of quiescent phase 
and exhibited a flux peak with $\sim10^{-9}$ erg cm$^{-2}$ s$^{-1}$ around January 9--10 (see figure 2 in \cite{Serino15}). 
Then the intensity gradually decreased from January 10 to January 13, 
and reached a mean flux of $\sim10^{-10}$ erg cm$^{-2}$ s$^{-1}$ in the period of January 14 to January 19. 
Thus, we assume that light from the outburst phase made the annular emission: 
the outer region of the annular emission is occupied by the X-rays of earlier epoch, while the inner region is by the later epoch.
Here, we also assume that the epoch of the flux peak on January 9 is corresponding to the brightest radius of 6$'$. 
The Suzaku observations were carried out about 21--24 days after the epoch of the flux peak, 
and hence the difference in the light traveling times between the direct and the dust scattered lights would be about 21--24 days. 

As is shown in section 3, MAXI\,J1421$-$613 is located behind the 43~km~s$^{-1}$ CO molecular cloud. The LSR of $\sim43$~km~s$^{-1}$ gives two solutions for the distance of the molecular cloud of  $\sim2.6$~kpc and $\sim9$~kpc. 
The distance to MAXI J1421-613 can be calculated using the difference in the light traveling times, the distance to the dust layer, and the width of the annular emission (equation 8 in Tr\"{u}mper \& Sch\"{o}nfelder 1973). 
 We assume that the dust layer responsible for the annular emission is located at the same position as the CO cloud, namely 2.6~kpc or 9~kpc. If the dust layer is located at 2.6~kpc from the Sun, the distance to MAXI\,J1421$-$613 is calculated to be 3~kpc. In the case of 9~kpc, the distance  to MAXI\,J1421$-$613 is calculated to be 30~kpc, which is beyond the Galaxy, and thus unlikely.  

 We propose that the distance to MAXI J1421$-$613 is 3 kpc on the assumption that the dust layer lies at the same position as the  43~km~s$^{-1}$ CO molecular cloud. 
This demonstrates that they are located at the Scutum-Centaurus arm in the Galaxy. 
In the cases, the radius expansion of the annular emission is consistent with the value during the Suzaku observations ($\sim$30$''$ per 3 days, see section 3).

\section{Conclusion}\label{sec:conc}
We note that the preceding analysis with Suzaku by \citet{Serino15} failed to find the annular emission, and hence the present paper reports the new discovery of the annular emission. Our detection of the annular emission with no bright point source (figure~\ref{3band_image}) catches very rare chance of the dust scattering echo; MAXI\,J1421$-$613 showed an outburst phase (10$^{-9}$--10$^{-10}$~erg~cm$^{-2}$~s$^{-1}$) within a short duration of $<$ 10~days  whereas  the scattered X-rays are observed after MAXI\,J1421$-$613 re-entered to a long quiescent phase ($<10^{-13}$~erg~cm$^{-2}$~s$^{-1}$).
The steeper $\Gamma$ of the annular emission of $\sim4.2$ than that of MAXI\,J1421$-$613 of $\sim2.1$, and the time history of the radius and the flux is all consistent with a dust scattering echo. If MAXI\,J1421$-$613 shows an outburst again, one would be able to observe a dust scattering echo when they observe the object  more than ten days after the outburst.

\begin{ack}
We thank all the members of the Suzaku and MAXI teams. 
We are grateful to Dr. Katsuji Koyama and Dr. Yasuharu Sugawara for their valuable comments. 
K.K.N. is supported by Research Fellowships of JSPS for Young Scientists. This work was supported by JSPS and MEXT KAKENHI Grant Numbers JP16J00548 (KKN) and JP17K14289 (MN).
\end{ack}


\begin{thebibliography}{}
\bibitem[Bamba et al.(2003)]{Bamba2003}Bamba, A., Yamazaki, R., Ueno, M., \& Koyama, K. 2003, \apj, 589, 827
\bibitem[Bozzo et al.(2014)]{Bozzo14}Bozzo, E. et al. 2014, Astronomer's Telegram, 5765
\bibitem[Chakrabarty et al.(2014)]{Chakrabarty14}Chakrabarty, D., Jonker, P. G., \& Markwardt, C. B. 2014, Astronomer's Telegram, 5894
\bibitem[Dame et al.(2001)]{Dame01}Dame, T. M., Hartmann, D., \& Thaddeus, P. 2001,  ApJ, 547, 792
\bibitem[Draine(2003)]{Draine03}Draine, B. T. 2003, ApJ, 598, 1026
\bibitem[Heinz et al.(2015)]{Heinz15}Heinz, S. et al. 2015, ApJ, 806, 265
\bibitem[Heinz et al.(2016)]{Heinz16}Heinz, S., Corrales, L., Smith, R., Brandt, W. N., Jonker, P. G., Plotkin, R. M., \& Neilsen, J. 2016, ApJ, 825, 15
\bibitem[Ishisaki et al.(2007)]{Ishisaki07}Ishisaki, Y.  et al. \ 2007, \pasj, 59, S113
\bibitem[Koyama et al.(1997)]{Koyama1997}Koyama, K., Kinugasa, K., Matsuzaki, K., Nishiuchi, M., Sugizaki, M., Torii, K., Yamauchi, S., \& Aschenbach, B. 1997, PASJ, 49, L7
\bibitem[Koyama et al.(2007)]{Koyama07}Koyama, K., Tsunemi, H., Dotani, T., et al. \ 2007, PASJ, 59, S23
\bibitem[Mao et al.(2014)]{Mao14}Mao, J., Ling, Z., \& Zhang, S.-N. 2014, ApJ, 785, 23
\bibitem[McClure-Griffiths \& Dickey(2007)]{McClure07}McClure-Griffiths, N. M., \& Dickey, J. M. 2007, ApJ, 671, 427
\bibitem[Mitsuda et al.(1984)]{Mitsuda84}Mitsuda, K. et al. \ 1984, PASJ, 36, 741
\bibitem[Mitsuda et al.(2007)]{Mitsuda07}Mitsuda, K. et al. \ 2007, PASJ, 59, S1
\bibitem[Morooka et al.(2014)]{Morooka14}Morooka, Y., et al. 2014,  Astronomer's Telegram, 5750
\bibitem[Negoro et al.(2010)]{Negoro10}Negoro, H., et al. 2010, in ASP Conf. Ser., 434, Astronomical Data Analysis Software and Systems XIX, ed. Y. Mizumoto et al. (San Francisco: ASP), 127
\bibitem[Overbeck(1965)]{Overbeck65}Overbeck, J. W. 1965, ApJ, 141, 864
\bibitem[Pintore et al.(2017)]{Pintore17}Pintore, F., et al. 2017, MNRAS, 472, 1465
\bibitem[Predehl \& Schmitt(1995)]{Predehl95} Predehl, P., \& Schmitt, J. H. M. M. 1995, A\&A, 293, 889
\bibitem[Predehl et al.(2000)]{Predehl00}Predehl, P., Burwitz, V., Paerels, F., \& Tr\"umper, J. 2000, A\&A, 357, L25
\bibitem[Sasaki et al.(2004)]{Sasaki04}Sasaki, M., Plucinsky, P.~P., Gaetz, T.~J., Smith, R.~K., Edgar, R.~J., Slane, P.~O. \ 2004, \apj, 617, 322 
\bibitem[Svirski et al.(2011)]{Svirski11}Svirski, G., Nakar, E., \& Ofek, E. O. 2011, MNRAS, 415, 2485
\bibitem[Serino et al.(2015)]{Serino15} Serino, M. et al.  2015, \pasj, 67, 30
\bibitem[Serlemitsos et al.(2007)]{Serle07}Serlemitsos, P. J. et al. \ 2007, PASJ, 59, S9
\bibitem[Tawa et al.(2008)]{Tawa08}Tawa, N. et al. \ 2008,  PASJ,  60, S11
\bibitem[Tiengo \& Mereghetti(2006)]{Tiengo06}Tiengo, A., \& Mereghetti, S. 2006, A\&A, 449, 203
\bibitem[Tiengo et al.(2010)]{Tiengo10}Tiengo, A. et al.  2010, ApJ, 710, 227
\bibitem[Tr\"{u}mper \& Sch\"{o}nfelder(1973)]{Trumper73}Tr\"{u}mper, J \& Sch\"{o}nfelder, V, 1973, A\&A, 25, 445
\bibitem[Tuohy \& Garmire(1980)]{Tuohy80}Tuohy, I. and  Garmire, G. 1980,  ApJL, 239, L107
\bibitem[Uchiyama et al.(2008)]{Uchiyama08b}Uchiyama, Y., et al. 2008, PASJ, 60, S35
\bibitem[Vasilopoulos \& Petropulou(2016)]{Vasilo16}Vasilopoulos, G., \& Petropoulou, M. 2016, MNRAS, 455, 4426
\bibitem[Vaughan et al.(2004)]{Vaughan04}Vaughan, S.,  et al. 2004, ApJ, 603, L5
\bibitem[Vaughan et al.(2006)]{Vaughan06}Vaughan, S., et al. 2006, ApJ, 639, 323
\bibitem[Vianello et al.(2007)]{Vianello07}Vianello, G., Tiengo, A., \& Mereghetti, S. 2007, A\&A, 473, 423
\bibitem[Xiang et al.(2011)]{Xiang11}Xiang, J., Lee, J. C., Nowak, M. A. \& Wilms, J. 2011, ApJ, 738, 78
\end{thebibliography}
\end{document}